# Rotation Periods of Binary Asteroids with Large Separations – Confronting the Escaping Ejecta Binaries Model with Observations

D. Polishook [a,b,*], N. Brosch [b], D. Prialnik [a]

[a] Department of Geophysics and Planetary Sciences, Tel-Aviv University, Israel.

[b] The Wise Observatory and the School of Physics and Astronomy, Tel-Aviv University, Israel.

[*] Corresponding author: David Polishook, david@wise.tau.ac.il

**Abstract**

Durda et al. (2004), using numerical models, suggested that binary asteroids with large separation, called Escaping Ejecta Binaries (EEBs), can be created by fragments ejected from a disruptive impact event. It is thought that six binary asteroids recently discovered might be EEBs because of the high separation between their components (~100 > $a/R_p$ > ~20).

However, the rotation periods of four out of the six objects measured by our group and others and presented here show that these suspected EEBs have fast rotation rates of 2.5 to 4 hours. Because of the small size of the components of these binary asteroids, linked with this fast spinning, we conclude that the rotational-fission mechanism, which is a result of the thermal YORP effect, is the most likely formation scenario. Moreover, scaling the YORP effect for these objects shows that its timescale is shorter than the estimated ages of the three relevant Hirayama families hosting these binary asteroids. Therefore, only the largest (D~19 km) suspected asteroid, *(317) Roxane*, could be, in fact, the only known EEB.

In addition, our results confirm the triple nature of *(3749) Balam* by measuring mutual events on its lightcurve that match the orbital period of a nearby satellite in addition to its distant companion. Measurements of *(1509) Esclangona* at different apparitions show a unique shape of the lightcurve that might be explained by color variations.

# 1. Motivation and Background

As the number of discovered binary asteroids and systems of asteroids increased dramatically in the last decade, different physical mechanisms have been suggested to explain the formation of these objects. The rotational-fission of a small-sized (D<10 km) asteroid into two or more components (Scheeres 2007) is one of the most successful mechanisms to explain the existence of these binaries. Rotational-fission occurs after the asteroid is spun-up by the YORP effect: this is a thermal torque on a rotating irregular body active due to the asymmetric re-emission of sunlight (Rubincam 2000). These two mechanisms were observed: the YORP effect was directly measured on four asteroids (Lowry et al. 2007, Taylor et al. 2007, Kaasalenien et al. 2007, Durech et al. 2008, Durech et al. 2009) and recent photometric observations of asteroid "pairs" have showed that the rotational-fission mechanism can disrupt asteroids due to fast rotations (Pravec et al. 2010). This process takes place because asteroids are weak aggregate bodies with negligible tensile strength bound together only by gravity (Richardson et al. 2002). This structure, often referred to as a "rubble-pile" or as a shattered interior structure, determines a critical rotation period of about 2.2 hours; a faster rotation will cause the asteroid to disrupt. Therefore, primaries of binary asteroids and of asteroid pairs usually rotate fast, on order of 2 to 5 hours (Pravec et al. 2006; Pravec et al. 2010), which is a remnant to the crossing of their "self-disruptive speed limit".

However, not all binary asteroids can be explained as results of the rotational-fission mechanism. The YORP effect is probably not efficient enough to spin-up large-sized asteroids in the scale of D>10 km (Polishook and Brosch 2009) and Trans-Neptunian Objects (TNOs) are too distant from the Sun to develop a significant thermal torque. Since such systems obviously exist and were observed (Descamps et al. 2007, Marchis et al. 2005, Sheppard and Jewitt 2004), other mechanisms for binaries formation are needed.

One of these suggested mechanisms is based on collisions between main belt asteroids that result in Escaping Ejecta Binaries (EEBs). This mechanism, first described by Durda et al. 2004 (and based on older ideas by Hartmann 1979 and Durda 1996), acts on two asteroidal fragments ejected during a collision incident which become gravitationally bound together as they drift away from their parent-body. Using numerical models, Durda et al. (2004), showed that such scenarios are physically plausible and the resultant components are stable. In addition, such objects can have large values of physical parameters such as the rotation period of the primary, the size ratio between the components, and the separation between them. Moreover, the recent (2002 – 2009) observational discovery using adaptive optics, of six binary asteroids (Merline et al. 2009) with high component separation (~100 > a/Rp > ~20, while the rest of the binaries have values of about 10 and less), enabled Durda et al. (2010) to show how the physical parameters (see Table I), mainly the size-ratio and the components' separation (in units of the primaries radii) of these binaries, are similar to the numerical results of EEBs (see Fig. 3 at Durda et al. 2010).

While these items support the EEB forming mechanism, they cannot rule out other mechanisms, especially the YORP effect followed by the rotational-fission. Measuring the rotation periods of these asteroids can constrain the binary forming mechanism, and might indicate weather these six objects are YORP-binaries or Escaping Ejecta binaries, and if such EEBs are actually exist. Therefore, we performed a photometric campaign on these asteroids in order to derive their rotation periods.

## 2. Observations, Reduction, Measurements, Calibration and Analysis

Observations were performed using the two telescopes of the Wise Observatory (code: 097): the 1-m Ritchey-Chrétien telescope; and the 0.46-m Centurion telescope (see Brosch et al. 2008 for a description of the telescope and its performance). The 1-m telescope is equipped with a cryogenically-cooled Princeton Instruments (PI) CCD. At the f/7 focus of the telescope this CCD covers a field of view of 13'x13' with 1340x1300 pixels (0.58" per pixel, unbinned). Two CCDs were used at the prime focus of the 0.46-m while the research took place: a wide-field SBIG ST-10XME (40.5'x27.3' with 2184x1472 pixels, 1.1" per pixel, unbinned) used until mid-November 2008; and an SBIG STL-6303E with an even wider field of view (75'x55' with 3072x2048 pixels, 1.47" per pixel, unbinned) used afterwards. An R filter was used on the 1-m telescope while observations with the 0.46-m telescope were in "white light" with no filters ("Clear").

To achieve a point-like FWHM (at a seeing value of ~2.5 pixels), constrained by the observed angular velocity, exposure times of 60–180s seconds were chosen, all using an auto-guider. The observational circumstances are summarized in Table II, which lists the asteroid's designation, the observation date, the time span of the observation during that night, the number of obtained images, the object's heliocentric distance (r), geocentric distance (Δ), phase angle (α), and the Phase Angle Bisector (PAB) ecliptic coordinates ($L_{PAB}$, $B_{PAB}$ - see Harris et al. (1984) for the definition and ways of calculating these parameters). In addition, the object's mean observed magnitude is presented for nights when the measurements were calibrated with standard stars (see calibration method below).

The images were reduced in a standard way using bias and dark subtraction, and division by a normalized flatfield image. Times were corrected to mid-exposure. We used the IRAF *phot* function for the photometric measurements. Apertures with four-pixel radii were chosen to minimize photometric errors. The mean sky value was measured using an annulus of 10 pixels wide and inner radius 10 pixels around the asteroid.

After measuring, the photometric values were calibrated to a differential magnitude level using local comparison stars measured on every image using the same method as the asteroid. Variable stars were removed at a second calibration iteration leaving an average of ~650 local comparison stars per image (~30 with the PI CCD of the 1-m which has a much smaller field of view). The brightness of these stars remained constant to ±0.02 mag. A photometric shift was calculated for each image compared to a good reference image, using the local comparison stars. Some measurements were also calibrated to standard magnitude level by using the Landolt equatorial standards (Landolt 1992).

Astrometric solutions were obtained using *PinPoint* ([www.dc3.com](www.dc3.com)) and the asteroids were identified from the MPC web database. Analysis for the lightcurve period and amplitude was done using the Fourier series analysis (Harris and Lupishko 1989) based on periodic functions with two to ten harmonics. A best match of the observations to the Fourier series was found by least squares and is presented on each lightcurve. In case of *(3749) Balam*, where all the data was calibrated to standard magnitude, we also found the best fit for the *H-G* system. All results appear in Table III. See Polishook and Brosch 2008 (1-m) and Polishook and Brosch 2009 (0.46-m) for a complete description of the reduction, measurements, calibration and analysis.

*2.1 Analysis of Mutual Events*

The presence of a satellite around an asteroid can be derived from an asteroid's lightcurve when a mutual event is observed. This is seen as attenuation in the amount of light reflected from the asteroid during an occultation – when one component blocks the light reflected from the other component, or during an eclipse – when one component casts a shadow over its partner (Pravec et al. 2006). Therefore, any periodicity of the mutual events can be interpreted as the orbital period of the secondary around the primary. For this study two asteroids, *(1509) Esclangona* and *(3749) Balam*, were routinely observed to searched for mutual events, but only *(3749) Balam* presented such phenomena.

Since the light attenuations due to the mutual events are superposed on the lightcurve variability, which is in turn caused by the rotational period of the primary, the rotational periodicity must first be subtracted from the data so that the orbital periodicity can be derived. Therefore, the analysis for *(3749) Balam* also includes the following procedures:

1. Finding the rotation periodicity, as described above, using data without mutual events.
2. Producing a model with the derived rotational properties.
3. Subtracting the magnitude predicted by the model from all the observed data, including data with mutual events.
4. Searching for the period of the mutual events using Fourier series analysis (Harris and Lupishko 1989), with the same code used to derive the rotation period. Since the mutual events curve is characteristically flat with two attenuations, as many as 20 harmonics were used to describe its model.
5. Folding the data with the orbital period of the secondary. Such a lightcurve should present two mutual events (Pravec et al. 2006), usually a total event, when the attenuation reaches a plateau before increasing back to the average brightness, and a partial event, recognizable by its V-shape.

The properties of the mutual events curve can provide many insights about the characteristics of the asteroid system. A total eclipse/occultation seen in the lightcurve takes place when the light from the secondary is completely blocked by the primary or when the secondary is in front the primary blocking a fixed fraction of its surface. The depth of the total eclipse ($A_{mut}$) is a function of the two components' size ratio (*Rs/Rp*):

$$A_{mut} = 2.5 \log \left[ 1 + (\frac{R_s}{R_p})^2 \right] \quad (1)$$

When *Rs* is the radius of the secondary and *Rp* is the radius of the primary (Polishook and Brosch 2008, Descamps et al. 2008).

## 3. Results

We observed five out of the six objects suspected as EEBs: *(1509) Esclangona, (3749) Balam, (4674) Pauling, (17246) 2000 GL$_{74}$* and *(22899) 1999 TO$_{14}$*. The sixth, *(317) Roxane*, whose binary nature was discovered only recently (Merline et al. 2009) was previously observed photometrically by Harris et al. (1992) who found a rotation period of 8.1691±0.0003 hours for the asteroid's primary with an amplitude of 0.606±0.003 mag. The results for the other objects are described below and summarized in Table III.

*[put **Table III** about here]*

### *3.1. (1509) Esclangona*

*(1509) Esclangona* is an S-type asteroid (Tholen and Barucci 1989), belonging to the Hungaria group (a=1.866 AU). Its binary nature was discovered by Merline et al. (2003a) using direct imaging with ESO's VLT at a projected separation between the components of about 140 km (a/Rp = 23). A measurement with the Infra Red Astronomical Satellite (IRAS) approximates the primary diameter as 8.2±0.6 km (Tedesco et al. 2004), while less-secure, unresolved photometric observations, estimate it as 12 km (Merline et al. 2003b). Merline et al. 2003a measured a brightness difference of about 2.4 mag, suggesting that the companion has a diameter of ~2.8 km. Given *(1509) Esclangona's* binarity nature, Polishook et al. (2009) obtained rotationally resolved spectra to look for any variance in the colors of the asteroid as it rotates, and by that to learn if the satellite was disrupted from the primary and to constrain the system age. To obtain a complete picture of *(1509) Esclangona*, Polishook et al. performed photometric observations of the asteroid and derived a rotation period of 3.2524±0.0003 hours and amplitude of 0.16±0.01. Warner et al. (2010) observed *(1509) Esclangona* during September-October 2009 and derived two superposed periods of 3.25285±0.00002 hours and 6.6423±0.0003 hours with amplitudes of 0.13 mag and 0.04 mag, respectively. They suggested that the two periods are due to the rotations of the primary and the secondary.

Here we report on additional observations of *(1509) Esclangona* performed during 14 nights on April 2008 and August 2009 through February 2010. We derive a rotation period of 3.25250±0.00005 hours for the data from the 2009 apparition and amplitude of 0.17±0.02 mag (Fig. 1). We could not find any mutual events during all these nights, therefore, if *(1509) Esclangona* has a second close satellite, in addition to its already known and distant satellite, its orbital plane was not in the line of sight from Earth at the time of our observations. Future observations could derive the spin axis orientation of *(1509) Esclangona*, and might be able to completely reject the existence of a close binary.

In addition, the 2009 observations do present the unique features on the lightcurve seen on the 2008 apparition. These consist of *(i)* two peaks (phase 0 to 0.4) with varying amplitudes from 0.09 to nearly zero magnitude; *(ii)* followed by a wide minimum (phase 0.4 to 0.7) with small-scaled changing features; *(iii)* and a large amplitude peak (phase 0.7 to 1) of about 0.13 to 0.16 *mag*. While Warner et al. (2010) managed to remove some of these features by analyzing the lightcurve as two superposed periods (the rotation periods of the primary and the secondary), our data, taken on a wider range of observing nights, do not present any second periodicity (including their published periodicity of 6.6423±0.0003 hours), while the unique features on the lightcurve remain. We suggest that at least part of the unique shape of the lightcurve might be due to the color variations measured by Polishook et al. (2009). More observations at different apparitions and aspects could reveal the true shape of *(1509) Esclangona* and to determine the complete nature of its unique lightcurve.

*3.2. (3749) Balam*

Located in the main belt (a=2.237 AU) and part of the Flora family, this asteroid is actually a complex system that presents different types of asteroidal phenomena (discovered with different measuring methods). *(3749) Balam* has a close satellite orbiting only a 20 km away from the primary that was discovered from the presence of mutual events on its photometric lightcurve (Marchis et al. 2008a); a distant satellite located at a=289 km, observed by Merline et al. (2002) with AO on the Gemini-North telescope (the satellite's orbit was measured by Marchis et al. 2008b); and a secondary "pair", designated as *2009 BR$_{60}$*, found by numerical integrations (Vokrouhlický 2009) to have detached from *(3749) Balam* 150 to 500 kyr ago.

Here we report independent photometric measurements of *(3749) Balam* taken from July to December 2007. We derived a rotation period of 2.80490±0.00008 hours and an amplitude of 0.14±0.04 mag (Fig. 2), in agreement with the values published by Marchis et al. (2008a) of 2.80483±0.00002 hours and 0.13 mag respectively. We also present the phase curve of *(3749) Balam* from 7 to 27 degrees (Fig. 3). We derive an $H_R$ value of 13.20±0.02 mag and a G slope of 0.40±0.02. On December 13, 2007, *(3749) Balam* was observed in *BVRI* filters to derive its colors of: *B-V* = 0.84±0.01 mag, *V-R* = 0.46±0.01 mag and *R-I* = 0.37±0.01 mag. These colors match the taxonomy of S-type asteroids (Tholen and Barucci 1989) that have an averaged albedo of $P_V$ = 0.15. Recent spectroscopic observations of *(3749) Balam* in near infrared (Marchis et al. 2010) classified it in the similar Sq taxonomic class of DeMeo et al. (2009) that have the same albedo. By using this assumed albedo with the absolute magnitude $H_V$ = $H_R$ + *V-R* = 13.66 (±0.03) mag, the effective radius of *(3749) Balam* can be estimated as $R_{eff}$ = 3.18±0.05 km, using (Polishook and Brosch 2009):

$$R = \frac{1329}{2\sqrt{P_V}} 10^{-0.2 H_V} \quad (2)$$

In addition, some mutual events can be seen in our observations with a periodicity of 33.385±0.001 hours (Fig. 4); this is interpreted as the orbital period of the secondary about the primary. A total event with a magnitude of 0.18±0.02 mag, observed on 14 August 2007, was used to calculate the size ratio between the near satellite and the primary yielding Rs/Rp = 0.42±0.03. These values are also in agreement with Marchis et al. (2008a) who found an orbital period of 33.38±0.02 hours and Rs/Rp of 0.4±0.1.

*3.3. (4674) Pauling*

This Hungaria-group asteroid was found to be a binary by Merline et al. 2004 using AO at ESO's 8-m VLT. They estimated the satellite size as 2.5 km, assuming a primary with a diameter of 8 km, and a projected separation of about 250 km (a/Rp = 63).

We observed *(4674) Pauling* at November and December 2008 and derived a rotation period of 2.5307±0.0003 hours (Fig. 5), matching the value measured by Warner et al. (2006) of 2.5306±0.0003 hours. However, our derived amplitude of 0.14±0.04 mag is about twice the corresponding value of Warner et al. of 0.06±0.02 mag. This is due to observations at different phase angles between Warner et al.'s observations (at ~15°) to ours (at ~34°). A lack of telescope time did not allow us to broaden the photometric coverage to allow a search of the lightcurve for mutual events.

*3.4. (17246) 2000 GL$_{74}$*

Using HST images Tamblyn et al. (2004) discovered a satellite of *(17246) 2000 GL$_{74}$*, an outer main belt asteroid at a=2.839 AU that belongs to the Koronis family. The same group (Merline et al. 2009) estimated the diameter of the primary as 4.5 km and its secondary as about 1 km, with the projected distance between them of 228 km (a/Rp = 101).

We observed *(17246) 2000 GL$_{74}$* on 9 and 16 of December 2007 for 5.3 and 5.6 hours respectively, and could obtain a lightcurve with only a low S/N (Fig. 6). From these data it is unclear if only a segment from the full lightcurve has been observed (suggesting a rotation period larger than ~5 hours), or if an eclipse occurred during the observation time during December 9. *(17246) 2000 GL$_{74}$* will be back at opposition on September 2011, reaching an estimated V magnitude of 17.4 (see the MPC website), when we plan further observations.

*3.5. (22899) 1999 TO$_{14}$*

This Koronis-family asteroid in the outer main belt (a=2.845 AU) was recognized as a binary from a set of six HST images (Merline et al. 2003c). The diameter of the primary was estimated by Merline et al. (2009) as 4.5 km and its secondary is about 1 km, while the projected distance between them was 182 km (a/Rp = 81).

We observed *(22899) 1999 TO$_{14}$* on December 20, 2009 for 8.6 hours and derived a rotation period of 4.03±0.03 hours with an amplitude of 0.19±0.03 mag (Fig. 7). A previous observation with lower S/N at June 12, 2007 for 4.22 hours, supports this result. Unfortunately, a lack of telescope time did not allow us to broaden the photometric coverage to search for mutual events in the lightcurve.

**4. Discussion and Conclusions**

We presented here photometric observations and rotation periods for four of the six binary asteroids with relatively large (~100 > a/Rp > ~20) separations between their components, suspected to be EEBs by Durda et al. (2010) due to their large separations and their main belt orbits. Indeed, all these objects belong to Hirayama families: *(317) Roxane* and *(3749) Balam* are part of the Flora family; *(17246) 2000 GL$_{74}$* and *(22899) 1999 TO$_{14}$* are part of the Koronis family; and *(1509) Esclangona* and *(4674) Pauling* are within the Hungaria group, where a large number of asteroids from this group were formed after a catastrophic collision (Warner et al. 2009b). The dynamical classification of these asteroids is from the Asteroid Lightcurve Data Base (LCDB - Warner et al. 2009a).

Durda et al. (2010) do not discuss specifically the spin properties of the EEBs, although some additional spin rate should be present in EEBs formed by oblique impacts resulting presumably in slow spins, and commensurate with the orbital periods of the collided objects. However, the measured rotation periods of four of these objects, *(1509) Esclangona, (3749) Balam, (4674) Pauling* and *(22899) 1999 TO$_{14}$*, are quite fast, ranging from 2.5 to 4 hours (the fifth object, *(17246) 2000 GL$_{74}$*, has an unknown rotation period). Such spins match the theory of rotational-fission of a small-sized (D<10 km) asteroid: after the asteroid rotation rate increased due to the thermal YORP effect, the centrifugal force overcomes the weak gravity between the components of the rubble pile structure and the object disintegrate into two or more components (Scheeres 2007). All five asteroids have small diameters of 4.5 to 8.2 km; such bodies can be spun-up efficiently by the YORP effect and cross the rotation barrier for a rubble pile.

To estimate the timescale of the YORP effect on these asteroids, we scaled from timescales of asteroids demonstrated to have been spun-up by YORP. The rotation periods and spin acceleration rates of the following asteroids: (1620) Geographos (Durech et al. 2008), (1862) Apollo (Kaasalenien et al. 2007), (3103) Eger, (Durech et al. 2009) and (54509) YORP (Lowry et al. 2007 and Taylor et al. 2007) were taken from the relevant papers, while their diameters were taken from the asteroid LCDB (Warner et al. 2009a). Better estimation of 54509 YORP's diameter of 114 m appears in Lowry et al. (2007), based on radar observations. The YORP spin acceleration rates ($d\omega/dt$) were scaled by the semi-major axis (*a*) and the diameter (*D*) of each object (Table I) following the theory that $d\omega/dt$ from YORP decreases with increasing $a^2 \cdot D^2$. We derived a range of timescales, from ~115 to ~175 Myr, for the asteroids discussed here to double their spin rate. This excludes the sixth suspected EEB, *(317) Roxane*, that is too big to be affected by YORP. Such timescales are significantly shorter than the estimated ages of the three involved families: Flora – 0.5 to 1 Gyr (Nesvorný et al. 2002), Hungaria – ~0.5 Gyr (Warner et al. 2009b) and Koronis – 2 to 3 Gyr (Vokrouhlický et al. 2003). Moreover, the timescales calculated here represent the time needed to double the spin rate of an asteroid, meaning they might have been disrupted a long time before, considering their current fast rotations. The fact that the YORP effect has been measured to align the spin axes of other objects from the Koronis family (Slivan 2002), adds strength to the idea that the YORP effect had enough time to efficiently spin-up the rotation periods of *(17246) 2000 GL$_{74}$* and *(22899) 1999 TO$_{14}$* that could cause their fission. In addition, if these binaries were created when their family parent-body was destroyed by a collision, the binary YORP (BYORP) mechanism (Ćuk and Burns 2005) that destabilizes and dissolves binaries by radiation effects would by now have destroyed those binaries. Even though the BYORP mechanism was not measured yet, different models suggest that its timescale is about $10^5$ years (Ćuk and Burns 2005, McMahon and Scheeres 2010), which is very short compared to the ages of these collisional families. Therefore, the original collisions that created these families were probably not the mechanisms that created the studied binaries.

The only asteroid from the suspected EEBs list that was probably not created by the YORP effect and the rotational-fission mechanism is *(317) Roxane*: it is too big (D=18.7 km) to be spun-up efficiently by the YORP effect and it rotates too slowly (8.1691±0.0003 hours; Harris et al. 1992) for any type of rotational fission to have taken place. Therefore, this might be the only real EEB among the studied list of asteroids.

The Escaping Ejecta Binaries formation model may still be the trigger for the creation of the studied binaries in the case of more recent collisions. However, such energetic collisions should have resulted in a Hirayama family of their own, and currently, such are unknown for these objects. Moreover, the fast rotation periods of the measured binaries pose another difficulty for the EEB model. An ejected rock with a fast rotation should originate from a collision with a relatively fast impact velocity. However, high velocity impacts will results in less EEBs since few components will have the same trajectory and drifting velocity that might result in gravitationally bound objects. This is best shown in Fig. 6 of Durda et al. 2004, where the number of EEBs decreases for high-energy impacts. To conclude, orbital, physical and rotational properties of five out of the six suspected EEBs support the idea that these asteroids were probably spun-up by the YORP effect and then torn apart by the rotational fission mechanism to become binary asteroids.

**Acknowledgement**

David Polishook is grateful for an *Ilan Ramon* doctoral scholarship from the Israel Space Agency (ISA) and the Ministry of Science. We thank anonymous reviewers for helping us improving the paper. We thank the Wise Observatory staff for their continuous support.

**Tables:**

**Table I:** Physical parameters of the six binaries with high separation (Durda et al. 2010): semi major axis (a), primary radius (Rp), ratio between the secondary and the primary radii (Rs/Rp), projected separation between the components in km ($a_s$) and in units of the primary radius (a/Rp).

| Asteroid name | a [AU] | Rp [+] [km] | Rs/Rp | $a_s$ [km] | a/Rp |
|---|---|---|---|---|---|
| *(317) Roxane* | 2.286 | 9.35 | 0.29 | 257 | 27 |
| *(1509) Esclangona* ^ | 1.866 | 4.1 | 0.33 | 140 | 23 |
| *(3749) Balam* * | 2.237 | 3.3 | 0.21 | 289 | 90 |
| | | | 0.42 | 20 | 6 |
| *(4674) Pauling* | 1.859 | 4 | 0.43 | 250 | 63 |
| *(17246) 2000 $GL_{74}$* | 2.839 | 2.25 | 0.22 | 228 | 101 |
| *(22899) 1999 $TO_{14}$* | 2.845 | 2.25 | 0.22 | 182 | 81 |

+ The Rp values given here (excluding *(1509) Esclangona* and *(3749) Balam*) were reported by Durda et al. 2010 and determined by a measurement of the absolute magnitude H and a guess of the albedo value using Eq. 2.

^ *(1509) Esclangona* Rp was determined by the IRAS satellite (Tedesco et al. 2004).

* *(3749) Balam* parameters were presented by Marchis et al. 2008b that measured a semi major axis for the distanced satellite of 289 km, and a primary radius of 3.3 km. However, this radius value is actually the effective radius of the primary in addition to the near satellite of *(3749) Balam*. Therefore the real radius of the primary is smaller, and the Rs/Rp and a/Rp values should be modified accordingly.

**Table II:** Observation circumstances: asteroid name, observation date, nightly time span of the specific observation, the number of images obtained (N), the object's heliocentric (r), and geocentric distances (Δ), the phase angle (α), the Phase Angle Bisector (PAB) ecliptic coordinates ($L_{PAB}$, $B_{PAB}$), and the average magnitude (*mean observed R*) only after standard calibration. The mean magnitudes of observations that were taken without filters and were not calibrated to standard magnitude system are not presented.

| Asteroid name | Date | Time span [hours] | N | r [AU] | Δ [AU] | α [Deg] | $L_{PAB}$ [Deg] | $B_{PAB}$ [Deg] | Mean R [Mag] |
|---|---|---|---|---|---|---|---|---|---|
| *(1509) Esclangona* | Jan 7, 2008 | 4.83 | 114 | 1.84 | 1.01 | 22.42 | 135.3 | -12.7 | 14.79 |
| | Jan 8, 2008 | 1.52 | 42 | 1.84 | 1.00 | 22.06 | 135.5 | -12.9 | 14.76 |
| | Jan 14, 2008 | 5.34 | 164 | 1.83 | 0.96 | 19.91 | 136.6 | -14.1 | 14.64 |
| | Apr 27, 2008 | 2.19 | 68 | 1.81 | 1.36 | 33.43 | 158.4 | -25.3 | - |
| | Apr 28, 2008 | 2.04 | 65 | 1.81 | 1.37 | 33.49 | 158.7 | -25.3 | - |
| | Aug 26, 2009 | 2.42 | 61 | 1.93 | 1.26 | 28.27 | 13.9 | 27.8 | - |
| | Aug 27, 2009 | 4.06 | 79 | 1.93 | 1.26 | 28.13 | 14.0 | 27.9 | - |

|  | Sep 2, 2009 | 2.53 | 36 | 1.93 | 1.22 | 27.12 | 15.4 | 28.3 | - |
|  | Sep 14, 2009 | 3.04 | 62 | 1.93 | 1.15 | 24.87 | 17.8 | 29.0 | - |
|  | Sep 23, 2009 | 5.89 | 97 | 1.93 | 1.11 | 23.15 | 19.4 | 29.4 | - |
|  | Oct 18, 2009 | 5.39 | 78 | 1.92 | 1.07 | 20.46 | 23.5 | 29.2 | - |
|  | Oct 25, 2009 | 8.24 | 104 | 1.92 | 1.08 | 20.71 | 24.6 | 28.9 | - |
|  | Nov 21, 2009 | 5.93 | 84 | 1.92 | 1.17 | 24.87 | 29.3 | 26.5 | - |
|  | Dec 15, 2009 | 5.49 | 85 | 1.92 | 1.34 | 28.91 | 34.5 | 23.3 | - |
|  | Dec 20, 2009 | 2.57 | 15 | 1.91 | 1.39 | 29.52 | 35.8 | 22.5 | - |
|  | Jan 15, 2010 | 3.77 | 42 | 1.91 | 1.63 | 31.04 | 43.2 | 18.3 | - |
|  | Feb 9, 2010 | 0.95 | 17 | 1.90 | 1.88 | 30.22 | 51.7 | 14.2 | - |
| *(3749) Balam* | Jul 15, 2007 | 3.53 | 35 | 2.37 | 1.87 | 24.26 | 352.6 | 4.4 | 17.43 |
|  | Jul 19, 2007 | 3.62 | 37 | 2.36 | 1.82 | 23.83 | 353.3 | 4.6 | 17.24 |
|  | Jul 22, 2007 | 1.26 | 12 | 2.36 | 1.78 | 23.45 | 353.8 | 4.7 | 17.17 |
|  | Jul 23, 2007 | 3.65 | 32 | 2.36 | 1.77 | 23.30 | 354.0 | 4.7 | 17.00 |
|  | Aug 10, 2007 | 5.87 | 64 | 2.34 | 1.57 | 19.60 | 356.6 | 5.2 | 16.76 |
|  | Aug 12, 2007 | 5.31 | 64 | 2.34 | 1.54 | 19.04 | 356.8 | 5.3 | - |
|  | Aug 14, 2007 | 6.06 | 73 | 2.34 | 1.53 | 18.46 | 357.1 | 5.3 | - |
|  | Aug 16, 2007 | 6.25 | 47 | 2.33 | 1.51 | 17.84 | 357.3 | 5.4 | 16.59 |
|  | Oct 4, 2007 | 7.82 | 86 | 2.28 | 1.30 | 6.78 | 1.8 | 6.4 | 15.96 |
|  | Oct 5, 2007 | 6.95 | 71 | 2.28 | 1.30 | 7.19 | 1.9 | 6.4 | 15.94 |
|  | Oct 6, 2007 | 7.11 | 75 | 2.28 | 1.30 | 7.65 | 2.0 | 6.4 | 15.97 |
|  | Nov 5, 2007 | 5.61 | 25 | 2.24 | 1.47 | 19.88 | 5.3 | 6.4 | 16.57 |
|  | Dec 13, 2007 | 4.27 | 14 | 2.19 | 1.85 | 26.53 | 12.3 | 6.0 | 17.18 |
| *(4674) Pauling* | Nov 2, 2008 | 2.76 | 62 | 1.74 | 1.33 | 34.66 | 104.9 | -14.1 | 17.18 |
|  | Nov 3, 2008 | 2.15 | 48 | 1.74 | 1.32 | 34.62 | 105.3 | -14.3 | 17.07 |
|  | Nov 29, 2008 | 1.91 | 34 | 1.73 | 1.10 | 32.01 | 113.9 | -19.0 | - |
|  | Dec 26, 2008 | 2.01 | 31 | 1.73 | 0.94 | 26.44 | 121.6 | -23.1 | - |
| *(17246) 2000 GL$_{74}$* | Dec 9, 2007 | 5.26 | 46 | 2.90 | 1.92 | 2.58 | 71.6 | 1.8 | 17.41 |
|  | Dec 16, 2007 | 5.62 | 54 | 2.90 | 1.94 | 5.36 | 71.8 | 1.9 | - |
| *(22899) 1999 TO$_{14}$* | Jun 12, 2007 | 4.22 | 35 | 2.92 | 2.00 | 10.14 | 237.3 | 3.3 | - |
|  | Dec 20, 2009 | 8.60 | 103 | 2.85 | 1.87 | 1.12 | 89.5 | -2.5 | - |

**Table III:** Analysis results: asteroid's name, period, reliability code, amplitude, absolute magnitude $H_R$, the slope parameter G and range of phase angles.

| Name | Period [hours] | Reliability code | Amplitude [mag] | $H_R$ [mag] | G | Phase angle range [deg] |
|---|---|---|---|---|---|---|
| *(1509) Esclangona* | 3.25250±0.00005 | 3 | 0.17±0.02 | - | - | 19.9 – 33.5 |
| *(3749) Balam* | 2.80490±0.00008 | 3 | 0.14±0.04 | 13.20±0.02 | 0.40±0.02 | 6.8 – 26.5 |
| *(4674) Pauling* | 2.5209±0.0004 | 3 | 0.14±0.04 | - | - | 27.4 – 34.7 |
| *(17246) 2000 GL$_{74}$* | - | 1 | >~ 0.15 | - | - | 2.6 |
| *(22899) 1999 TO$_{14}$* | 4.03±0.03 | 2 | 0.19±0.03 | - | - | 1.1 |
| *(317) Roxane** | 8.1691±0.0003 | 3 | 0.606±0.003 | 10.07±0.01 | 0.49±0.04 | 0.4 – 10.7 |

* Reference: Harris et al. 1992.

**Figures:**

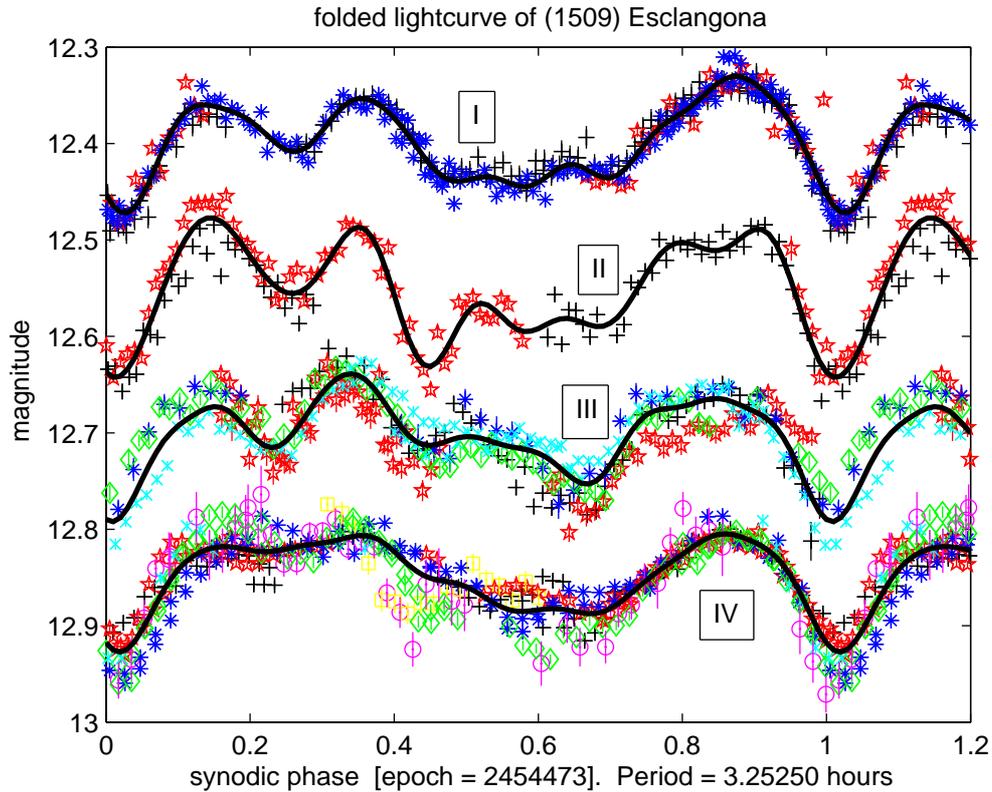

Fig. 1: *(1509) Esclangona* lightcurve, folded with a period of 3.25250 hours, which is the rotation period of the primary, with the resulted model (black line). The lightcurves were divided into four groups, by the apparition (2008 and 2009), before and after opposition. The top lightcurve includes data form January 2008 (I), and is calibrated to a standard magnitude scale. The three lightcurves below, representing data from April 2008 (II), August to September 2009 (III), and October 2009 through February 2010 (IV), were shifted on the Y-axis for display. Different symbols in each group represent different nights (see Table II for a summary of the observing nights). Because the 2009 observations were performed on different apparition than the 2008 observations, and the viewing angle changed dramatically between the apparitions, the 2009 lightcurves were shifted also on the X-axis so their features will match the 2008 lightcurves.

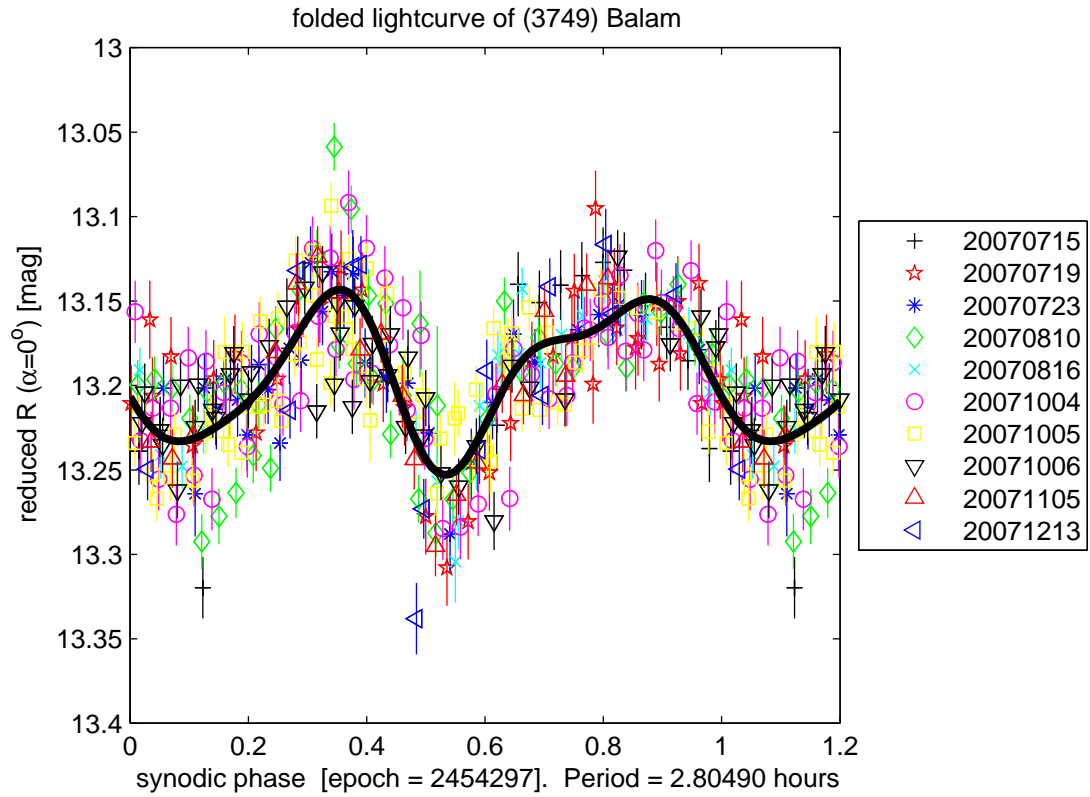

Fig. 2: *(3749) Balam* lightcurve, folded with a period of 2.80490 hours, which is the rotation period of the primary, with the resulted model (black line). Any mutual events were excluded from this lightcurve.

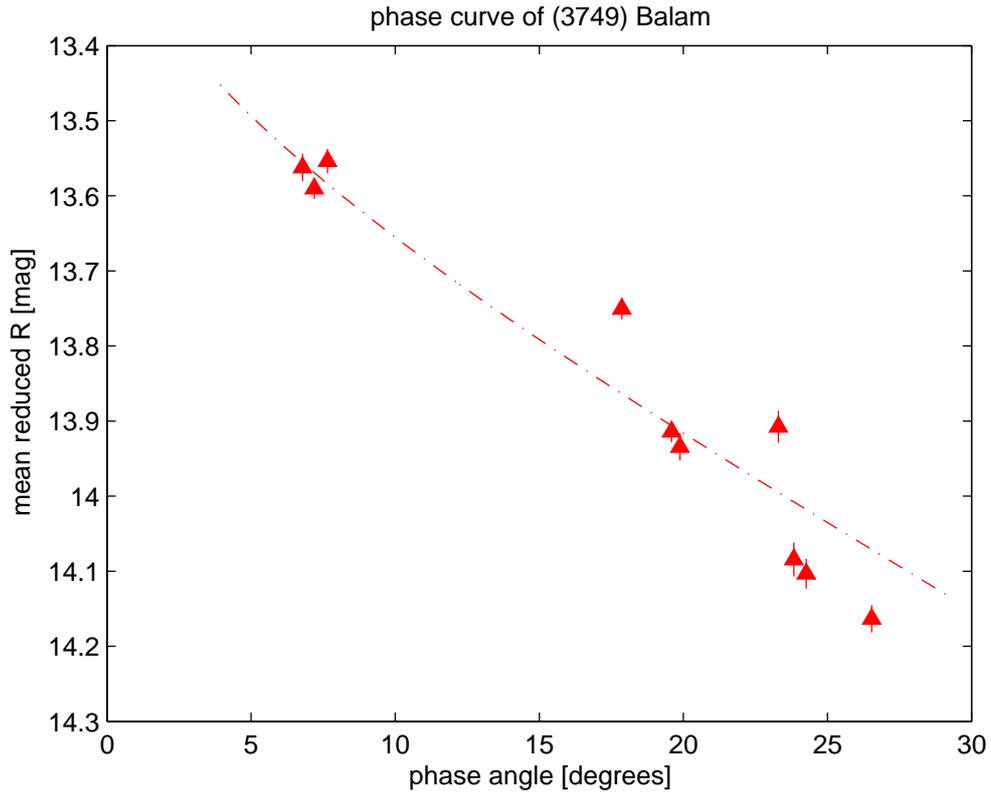

Fig. 3: The phase curve of *(3749) Balam* that match a $H_R$ value of 13.20±0.02 mag and a G slope of 0.40±0.02 (dashed line). Each triangular represents the mid-amplitude of the lightcurve at a specific night.

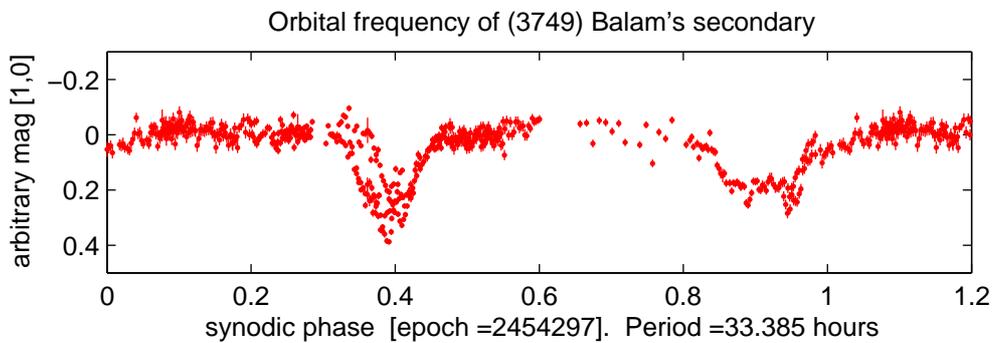

Fig. 4: Mutual events on the lightcurve of *(3749) Balam* after the rotation period of the primary was subtracted. The plot is folded with a period of 33.385±0.001 hours, which is the orbital period of the inner satellite of *(3749) Balam*.

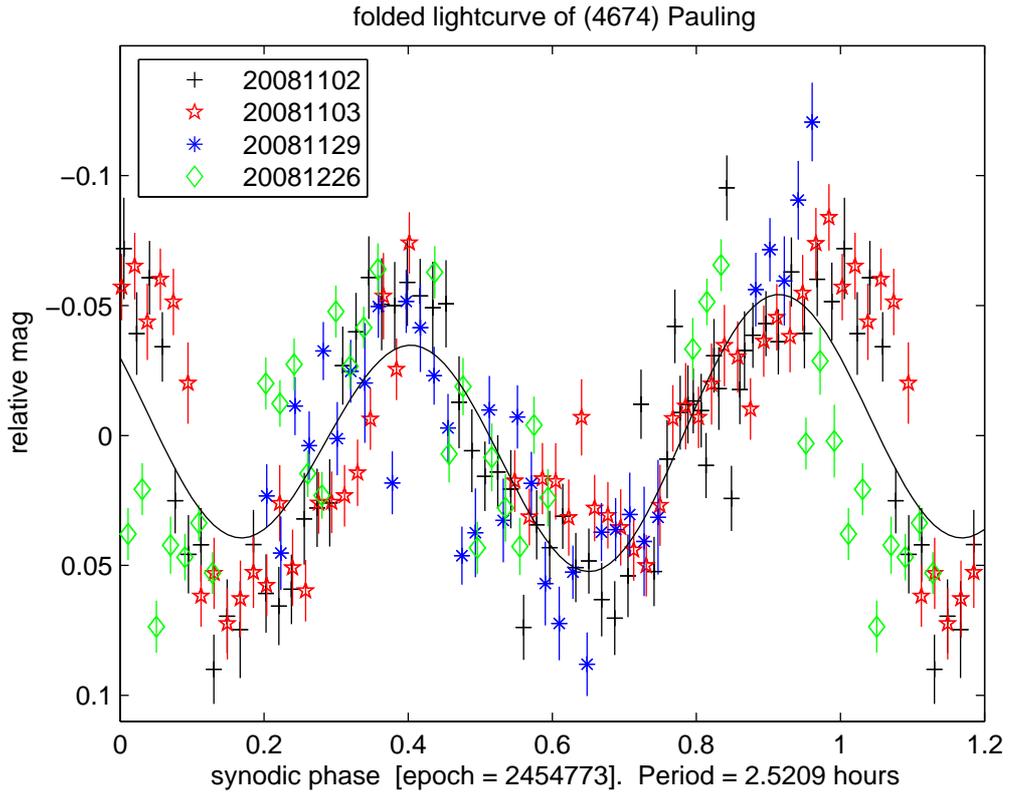

Fig. 5: *(4674) Pauling* lightcurve, folded with a period of 2.5209 hours, which is the rotation period of the primary, with the resulted model (black line).

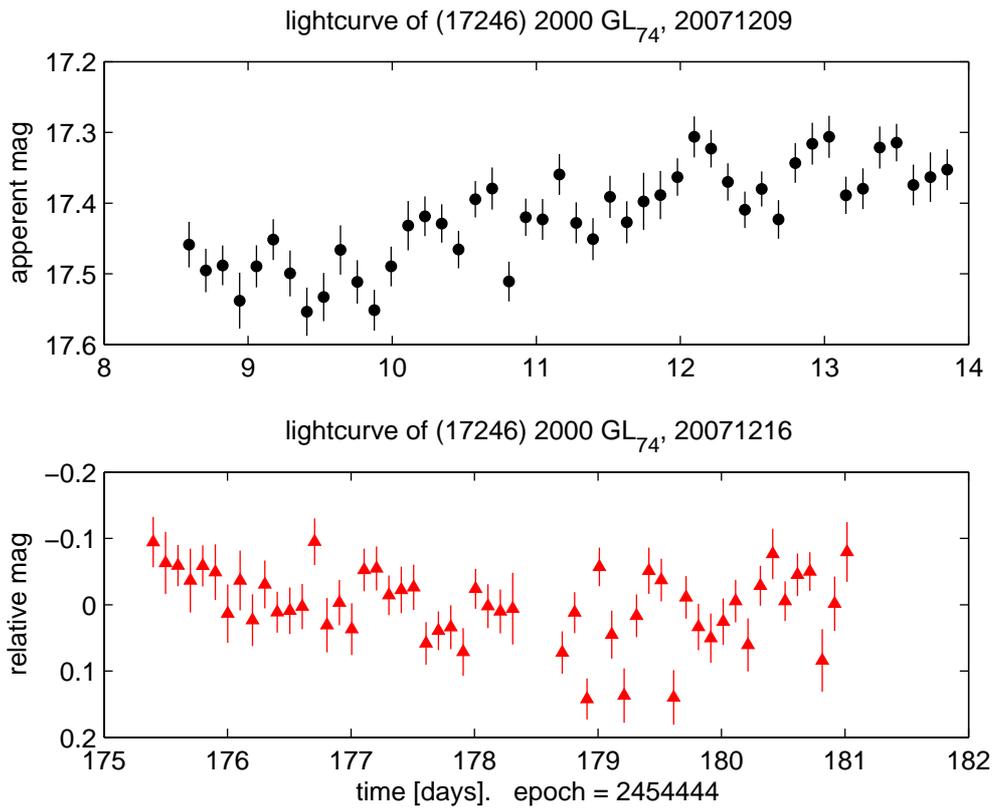

Fig. 6: *(17246) 2000 GL₇₄* lightcurves, from December 9$^{th}$ (top panel) and 16$^{th}$ (lower panel) 2007. We could not derive the rotation period of the asteroid using these poor S/N measurements.

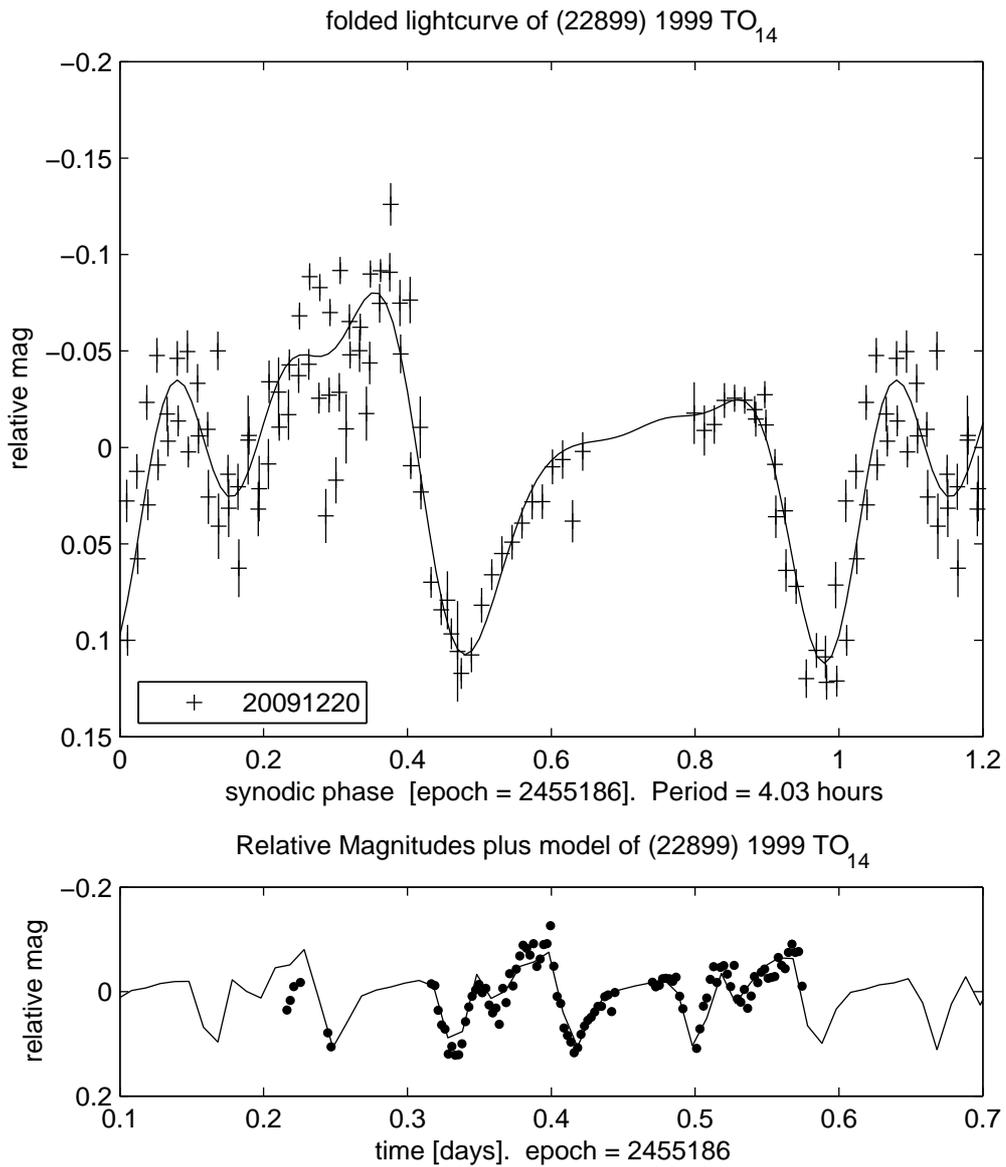

Fig. 7: *(22899) 1999 TO₁₄* lightcurve, folded with a period of 4.03 hours (top panel), which is the rotation period of the primary, with the resulted model (black line). The reduced data points are presented in the lower panel to better display the periodicity of the observed data.